\documentstyle[12pt]{article}
\begin{document}
\title{COLLAPSING SPHERES SATISFYING AN ``EUCLIDEAN CONDITION''}
\author{L. Herrera$^1$\thanks{e-mail: laherrera@cantv.net.ve} and N. O. Santos$^{2,3,4}$
\thanks{e-mail: N.O.Santos@qmul.ac.uk and nilton.santos@upmc.fr}\\
{\small $^1$Escuela de F\'{\i}sica, Facultad de Ciencias,}\\
{\small Universidad Central de Venezuela, Caracas, Venezuela.}\\
{\small $^2$School of Mathematical Sciences, Queen Mary,}\\
{\small University of London, London E1 4NS, UK.}\\
{\small $^3$Laborat\'orio Nacional de Computa\c{c}\~ao Cient\'{\i}fica,
25651-070 Petr\'opolis RJ, Brazil.}\\
{\small $^4$Universit\'e Pierre et Marie Curie, LERMA(ERGA) CNRS - UMR 8112,}\\
{\small 94200 Ivry, France.}}
\maketitle
\begin{abstract}
We study the general properties of fluid spheres satisfying the heuristic assumption that theirs  areal and   proper radius are equal (the Euclidean condition). Dissipative and non-dissipative models are considered. In the latter case, all models are necessarily geodesic and  a subclass of the Lema\^{\i}tre-Tolman-Bondi  solution is obtained. In the dissipative case solutions are non-geodesic and are characterized by the fact that all non-gravitational forces acting on any fluid element produces a radial three-acceleration independent on  its inertial mass.
\end{abstract}
\newpage
\section{Introduction}
Analytical or numerical solutions  to Einstein equations describing dissipative gravitational collapse are thought to be useful not only for describing specific astrophysical  phenomena,  but also as test-bed for probing cosmic  censorship and hoop conjecture among other important issues.

In this work we consider a large family of solutions, derived from the heuristic assumption that the areal radius of  any  shell of fluid, which is the radius obtained from its area, equals the proper radial distance from the centre to the shell. Since these two quantities are  always equal in the Euclidean geometry, systems described by solutions satisfying such a condition will be called ``Euclidean stars''.

Some of the  models are necessarily dissipative. This is appealing from a physical point of view, since  gravitational collapse is a
highly dissipative process (see \cite{Hs,Hetal,Mitra} and references
therein). This dissipation is required to account for the very large
(negative) binding energy of the resulting compact object of the order
of $-10^{53}$ erg.

The resulting dissipative models have a distinct dynamical property, namely all non-gravitational forces acting on any fluid element, produce a  radial three-acceleration being independent of the inertial mass density of the fluid element. This  behaviour, which is characteristic of the gravitational force, is now shared, due to the Euclidean condition, by all forces. The specific case of the shear-free and conformally flat fluid, are considered in detail.

Non-dissipative models are necessarily geodesic, belonging to the Lema\^{\i}tre-Tolman-Bondi (LTB) solutions (more specifically to the parabolic subclass). They may describe collapsing dust or, more generally, anisotropic fluids \cite{Herrera}.

\section{The Euclidean condition and its consequences}
We consider a spherically symmetric distribution  of collapsing
fluid, bounded by a spherical surface $\Sigma$. The fluid is
assumed to be locally anisotropic with principal stresses unequal and undergoing dissipation in the
form of heat flow.
Choosing comoving coordinates inside $\Sigma$, the general
interior metric can be written
\begin{equation}
ds^2_-=-A^2dt^2+B^2dr^2+R^2(d\theta^2+\sin^2\theta d\phi^2),
\label{1}
\end{equation}
where $A$, $B$ and $R$ are functions of $t$ and $r$ and are assumed
positive. We number the coordinates $x^0=t$, $x^1=r$, $x^2=\theta$
and $x^3=\phi$.

The matter energy-momentum $T_{\alpha\beta}^-$ inside $\Sigma$
has the form
\begin{equation}
T_{\alpha\beta}^-=(\mu +
P_{\perp})V_{\alpha}V_{\beta}+P_{\perp}g_{\alpha\beta}+(P_r-P_{\perp})\chi_{
\alpha}\chi_{\beta}+q_{\alpha}V_{\beta}+V_{\alpha}q_{\beta}, \label{3}
\end{equation}
where $\mu$ is the energy density, $P_r$ the radial pressure,
$P_{\perp}$ the tangential pressure, $q^{\alpha}$ the heat flux,
$V^{\alpha}$ the four-velocity of the fluid and
$\chi^{\alpha}$ a unit four-vector along the radial direction.
These quantities satisfy
\begin{equation}
V^{\alpha}V_{\alpha}=-1, \;\; V^{\alpha}q_{\alpha}=0, \;\; \chi^{\alpha}\chi_{\alpha}=1, \;\;
\chi^{\alpha}V_{\alpha}=0.
\end{equation}
The four-acceleration $a_{\alpha}$ and the expansion $\Theta$ of the fluid are
given by
\begin{equation}
a_{\alpha}=V_{\alpha ;\beta}V^{\beta}, \;\;
\Theta={V^{\alpha}}_{;\alpha}, \label{4b}
\end{equation}
and its  shear $\sigma_{\alpha\beta}$ by
\begin{equation}
\sigma_{\alpha\beta}=V_{(\alpha
;\beta)}+a_{(\alpha}V_{\beta)}-\frac{1}{3}\Theta(g_{\alpha\beta}+V_{\alpha}V
_{\beta}).
\label{4a}
\end{equation}
Since we assumed the metric (\ref{1}) comoving then
\begin{equation}
V^{\alpha}=A^{-1}\delta_0^{\alpha}, \;\;
q^{\alpha}=qB^{-1}\delta^{\alpha}_1, \;\;
\chi^{\alpha}=B^{-1}\delta^{\alpha}_1, \label{5}
\end{equation}
where $q$ is a function of $t$ and $r$.
From  (\ref{4b}) with (\ref{5}) we have for the  four-acceleration and its scalar $a$,
\begin{equation}
a_1=\frac{A^{\prime}}{A}, \;\; a^2=a^{\alpha}a_{\alpha}=\left(\frac{A^{\prime}}{AB}\right)^2, \label{5c}
\end{equation}
and for the expansion
\begin{equation}
\Theta=\frac{1}{A}\left(\frac{\dot{B}}{B}+2\frac{\dot{R}}{R}\right),
\label{5c1}
\end{equation}
where the  prime stands for $r$
differentiation and the dot stands for differentiation with respect to $t$.
With (\ref{5}) we obtain
for the shear (\ref{4a}) its non zero components
\begin{equation}
\sigma_{11}=\frac{2}{3}B^2\sigma, \;\;
\sigma_{22}=\frac{\sigma_{33}}{\sin^2\theta}=-\frac{1}{3}R^2\sigma,
 \label{5a}
\end{equation}
and its scalar
\begin{equation}
\sigma^{\alpha\beta}\sigma_{\alpha\beta}=\frac{2}{3}\sigma^2,
\label{5b}
\end{equation}
where
\begin{equation}
\sigma=\frac{1}{A}\left(\frac{\dot{B}}{B}-\frac{\dot{R}}{R}\right).\label{5b1}
\end{equation}
The mass function $m(t,r)$ introduced by Misner and Sharp
\cite{Misner} (see also \cite{Cahill}) reads
\begin{equation}
m=\frac{R^3}{2}{R_{23}}^{23}
=\frac{R}{2}\left[\left(\frac{\dot R}{A}\right)^2-\left(\frac{R^{\prime}}{B}\right)^2+1\right],
 \label{17masa}
\end{equation}

 We can define the velocity $U$ of the collapsing
fluid  as the variation of the areal radius with respect to proper time, i.e.\
\begin{equation}
U=D_TR<0 \;\; \mbox{(in the case of collapse)}, \label{19n}
\end{equation}
where $D_T=(1/A)(\partial/\partial t)$ defines the derivative with respect to proper time.
Then (\ref{17masa}) can be rewritten as
\begin{equation}
E \equiv \frac{R^{\prime}}{B}=\left(1+U^2-\frac{2m}{R}\right)^{1/2}.
\label{20x}
\end{equation}

The proper radial three-acceleration $D_TU$ of an infalling particle inside $\Sigma$ can
be calculated to obtain
\begin{equation}
D_TU=-\left(\frac{m}{R^2}+4\pi P_r R\right)
+Ea, \label{28n}
\end{equation}
feeding back this expression into the radial component of the Bianchi identities produces (see \cite{matter1} for details)
\begin{eqnarray}
\left(\mu+P_r \right)D_TU=-(\mu+P_r)\left(\frac{m}{R^2}+4\pi P_r R\right) \nonumber\\
-E^2\left[D_R P_r
+2(P_r-P_{\perp})\frac{1}{R}\right]
-E\left[D_T q+2q\left(2\frac{U}{R}+\sigma\right)\right].
\label{3m}
\end{eqnarray}
The  physical meaning of different terms in (\ref{3m}) is discussed in detail in \cite{Hs,matter1,matter2}. We would like just to recall that   the first term on the right hand side describes the gravitational force term.  As expected from the equivalence principle, its contribution to $D_TU$ is independent  on the inertial mass density $\mu+P_r$.
The two last terms describe non-gravitational force terms (i.e. their combination vanishes in a geodesic motion).

Two radii are determined for a collapsing spherical fluid distribution by the metric (\ref{1}). The first is determined by $R(t,r)$ representing the radius as measured by its spherical surface, hence called its {\it areal radius}. The second is obtained out its radial integration $\int B(t,r)dr$, hence called {\it proper radius}. These two radii in general, in Einstein's theory, need not to be equal, unlike in Newton's theory.
Here we assume  those two radii to be  equal. Hence with this condition we can write,
\begin{equation}
B=R^{\prime}, \label{16}
\end{equation}
implying from (\ref{20x})
\begin {equation}
E=1.\label{21E}
\end{equation}

The field equations with this condition  become
\begin{eqnarray}
\kappa\mu =\frac{1}{A^2}\left(\frac{\dot R}{R}+2\frac{{\dot R}^{\prime}}{R^{\prime}}\right)\frac{\dot R}{R}, \label{17}\\
\kappa qAR^{\prime}=-2\frac{\dot R}{R}\frac{A^{\prime}}{A}, \label{18}\\
\kappa P_r=-\frac{1}{A^2}\left[2\frac{\ddot R}{R}-\left(2\frac{\dot A}{A}-\frac{\dot R}{R}\right)
\frac{\dot R}{R}\right]+2\frac{A^{\prime}}{A}\frac{1}{RR^{\prime}}, \label{19}\\
\kappa P_{\perp}=-\frac{1}{A^2}\left[\frac{\ddot R}{R}+\frac{{\ddot R}^{\prime}}{R^{\prime}}
-\frac{\dot A}{A}\frac{\dot R}{R}
-\left(\frac{\dot A}{A}
-\frac{\dot R}{R}\right)\frac{{\dot R}^{\prime}}{R^{\prime}}
\right] \nonumber\\
+\left[\frac{A^{\prime\prime}}{A}-\left(\frac{R^{\prime\prime}}{R^{\prime}}
-\frac{R^{\prime}}{R}\right)\frac{A^{\prime}}{A}\right]\frac{1}{R^{\prime 2}}; \label{20}
\end{eqnarray}
while the mass function (\ref{17masa}) now reads,
\begin {equation}
m=\frac{R}{2}\left(\frac{\dot R}{A}\right)^2. \label{21}
\end{equation}

It is clear from (\ref{21}) that if ${\dot R}=0$ then $m=0$ and spacetime becomes Minkowskian. Therefore {\it all Euclidean stars are necessarily non-static}. Furthermore, using  (\ref{19n}),
 (\ref{21}) can be rewritten as
\begin{equation}
\frac{m}{R}=\frac{U^2}{2}. \label{24}
\end{equation}
Hence, (\ref{24}) can be interpreted as the Newtonian kinetic energy (per unit mass) of the collapsing particles being equal to their Newtonian potential energy.

From (\ref{18}), we observe that if the system is dissipating   in the form of heat flow, the collapsing source needs $A^{\prime}\neq 0$, implying because of (\ref{5c}) $a^{\alpha}\neq 0$. This means that dissipation does not allow collapsing particles to follow geodesics. Inversely, of course, {\it non-dissipative Euclidean models are necessarily geodesic}, since $q=0$ implies because of (\ref{5c}) and   (\ref{18}) that $a^\alpha=0 $.

It is interesting to observe that due to the Euclidean condition, the dynamical equation (\ref{28n}) or (\ref{3m}) becomes,
\begin{equation}
D_TU=-\left(\frac{m}{R^2}+4\pi P_r R\right)-\frac{\kappa qR}{2U},
\label{3mnubis}
\end{equation}
implying that the non-gravitational force term (the last on the right hand side) contributes to $D_tU$, for any fluid element, independently on its inertial mass density. In other words, the Euclidean condition produces a ``gravitational-like''  behaviour in  non-gravitational forces (which are controlled by $q$).
Thus, the effect of non-gravitational forces amounts to modify the gravitational force term, leaving a ``gravitational-like'' force term producing a radial three-acceleration independent on  the inertial mass density of the fluid element.

The Weyl tensor $C_{\alpha\beta\gamma\delta}$ for metric (\ref{1}) with (\ref{16}) has the following non zero components,
\begin{eqnarray}
C_{0101}=\frac{A^2}{3}\left\{\left[\frac{\ddot R}{R}-\frac{{\ddot R}^{\prime}}{R^{\prime}}+
\left(\frac{\dot A}{A}+\frac{\dot R}{R}\right)\left(\frac{{\dot R}^{\prime}}{R^{\prime}}-\frac{\dot R}{R}\right)\right]\left(\frac{R^{\prime}}{A}\right)^2\right. \nonumber\\
\left.+\frac{A^{\prime\prime}}{A}-\left(\frac{R^{\prime\prime}}{R^{\prime}}
+\frac{R^{\prime}}{R}\right)\frac{A^{\prime}}{A}\right\}, \label{24a}
\end{eqnarray}
and all the other non zero components are proportional to (\ref{24a}),
\begin{eqnarray}
\frac{R^2}{2}\;C_{0101}=-B^2C_{0202}=-\left(\frac{B}{\sin\theta}\right)^2C_{0303} \nonumber\\
=A^2C_{1212}=\left(\frac{A}{\sin\theta}\right)^2C_{1313}=-\frac{1}{2}\left(\frac{AB}{R\sin\theta}\right)^2C_{2323}. \label{24b}
\end{eqnarray}
With (\ref{5b1}), (\ref{19}) and (\ref{20}) we can rewrite (\ref{24a}) like
\begin{equation}
C_{0101}=\frac{AR^{\prime 2}}{3}\left[\kappa(P_{\perp}-P_r)A+2\frac{\dot R}{r}\sigma\right], \label{24ab}
\end{equation}
showing that for isotropic systems the shear-free conditions implies a conformally flat source.

We consider next  the non-dissipative case.

\section{Collapse with $q=0$}
As mentioned before, for this case we have from (\ref{18}) that $A^{\prime}=0$ which means $A=A(t)$ and by rescaling $t$ we can have
\begin{equation}
A=1.\label{25}
\end{equation}
Of course such models are members of  the Lema\^{\i}tre-Tolman-Bondi (LTB) spacetimes \cite{lemaitre,tolman,bondi}, furthermore they correspond to the parabolic case.
Indeed, the general metric  for LTB spacetimes read,
\begin{equation}
ds^2=-dt^2+\frac{R^{\prime 2}}{1-K(r)}dr^2+R^2(d\theta^2+\sin^2\theta d\phi^2),
\label{25IIIppnj}
\end{equation}
where $K(r)$ is an arbitrary function of $r$.

Imposing  the Euclidean condition (\ref{16}) in (\ref{25IIIppnj}), one obtains  $K=0$, which defines  parabolic LTB spacetimes. Further, assuming that the source consists of pure dust ($P_r=P_{\perp}=0$)
then it follows from the field equations that
\begin{equation}
R(t,r)=[c_1(r)t+c_2(r)]^{2/3}, \label{31}
\end{equation}
and
\begin{equation}
\kappa\mu=\frac{4c_1c_1^{\prime}}{3(c_1t+c_2)(c_1^{\prime}t+c_2^{\prime})}, \label{33}
\end{equation}
where $c_1(r)$ and $c_2(r)$ are integration functions.
Hence the solution reduces to  parabolic LTB collapsing dust \cite{lemaitre,tolman,bondi}.

From  (\ref{25}) and (\ref{31}) we have for (\ref{24a}),
\begin{equation}
C_{0101}=\left(\frac{2}{3}\right)^4\frac{c_1(c_1^{\prime}t+c_2^{\prime})^2}{(c_1t+c_2)^{5/3}}\;\sigma, \label{34a}
\end{equation}
where $\sigma$, from (\ref{5b1}), is
\begin{equation}
\sigma=\frac{c_2c_1^{\prime}-c_1c_2^{\prime}}{(c_1t+c_2)(c_1^{\prime}t+c_2^{\prime})}.
\end{equation}

In the shear-free case, $c_1=c_2$, the system becomes conformally flat too, and with the freedom for choosing the $r$ coordinate we can assume $c_1=r^{3/2}$ recovering the Friedmann critical dust sphere.

Of course more general models can be obtained by relaxing the condition of vanishing pressure, we recall that  LTB spacetime
is compatible with an anisotropic fluid \cite{gair,sussman}.

\section{Collapse with $q\neq 0$}
We consider now the dissipative case. For simplicity we assume the fluid to be shear-free. In this latter case the line element can be written as  \cite{Glass}
\begin{equation}
ds^2_-=-A^2dt^2+B^2[dr^2+r^2(d\theta^2+\sin^2\theta d\phi^2)], \label{35}
\end{equation}
then the Euclidean condition becomes
\begin{equation}
B=(Br)^{\prime}\rightarrow B=f(t)
\label{eu}
\end{equation}
implying
\begin{equation}
R=f(t)r, \label{36}
\end{equation}
where $f$ is an arbitrary function of $t$.

The field equations (\ref{17}-\ref{20}) now read,
\begin{eqnarray}
\kappa\mu =\frac{3}{A^2}\left(\frac{\dot f}{f}\right)^2, \label{37}\\
\kappa q =-2\frac{\dot f}{f^2}\frac{A^{\prime}}{A^2}, \label{38}\\
\kappa P_r=-\frac{1}{A^2}\left[2\frac{\ddot f}{f}-\left(2\frac{\dot A}{A}
-\frac{\dot f}{f}\right)\frac{\dot f}{f}\right]+\frac{2}{f^2r}\frac{A^{\prime}}{A}, \label{39}\\
\kappa P_{\perp}=-\frac{1}{A^2}\left[2\frac{\ddot f}{f}-\left(2\frac{\dot A}{A}
-\frac{\dot f}{f}\right)\frac{\dot f}{f}\right]+\frac{1}{f^2}\left(\frac{A^{\prime\prime}}{A}
+\frac{1}{r}\frac{A^{\prime}}{A}\right). \label{40}
\end{eqnarray}
From (\ref{39}) and (\ref{40}) we have
\begin{equation}
\kappa(P_{\perp}-P_r)=\frac{1}{f^2A}\left(A^{\prime\prime}-\frac{A^{\prime}}{r}\right). \label{41}
\end{equation}
From (\ref{24ab}) it follows that (in the shear--free case) if the collapsing source is conformally flat,  it must be isotropic in its pressures, and vice-versa.

The general form of all conformally flat and shear-free metrics is known \cite{hlsw}, it reads
\begin{equation}
A=\left[e_1\left(t\right)r^2+1\right]B, \label{II2}
\end{equation}
where $e_1$ is an arbitrary function of $t$,
and
\begin{equation}
B=\frac{1}{e_2(t)r^2+e_3(t)}, \label{II4}
\end{equation}
where $e_2$ and $e_3$ are arbitrary functions of $t$.

The Euclidean condition then implies
\begin{equation}
e_2=0, \;\; e_3=\frac{1}{f}.
\label{met}
\end{equation}

 An approximate solution of this kind has been presented and discussed in \cite{hlsw}. Furthermore,  an exact solution is also known \cite{GM}, which in turn is a particular  case of a family of solutions found in \cite{hja}. It reads (see Case III in \cite{GM})
\begin{equation}
f(t)=(\beta_1+\beta_2)^2 e^{-2\alpha r_{\Sigma} t}
\label{f}
\end{equation}
and
\begin{equation}
A=(\alpha r^2+1)f, \label{42}
\end{equation}
where $\alpha$, $\beta_1$ and  $\beta_2$ are constants.
The above solution satisfies junction conditions and its physical properties have been discussed in \cite{GM}. Thus we shall not  elaborate any further on it. Suffice to say at this point  that its physical properties are reasonable and a thermodynamic  analysis brings out the relevance of relaxational effects on the evolution of the system.

\section*{Acknowledgments.}
L.H. wishes to thank  financial support from the
FUNDACION EMPRESAS POLAR,  the  CDCH at Universidad Central
de Venezuela under grants PG 03-00-6497-2007 and PI 03-00-7096-2008, the  Universit\'e Pierre et Marie Curie (Paris) and Universitat  Illes Balears (Palma de Mallorca).

\end{document}